\documentclass[aps,prl,preprint,superscriptaddress]{revtex4-1}

\usepackage{url}
\usepackage{graphicx}
\usepackage{color}
\usepackage{dcolumn}
\usepackage{bm}

\newcommand\beq{\begin{equation}}
\newcommand\eeq{\end{equation}}

\begin{document}

\title{Reconfigurable anisotropy and functional transformations with VO$_{2}$-based metamaterial electric circuits}

\author{Salvatore Savo}
\email[]{savo@rowland.harvard.edu}
\affiliation{Rowland Institute at Harvard, Harvard University, Cambridge Massachusetts 02142, USA}

\author{You Zhou}
\affiliation{Harvard School of Engineering and Applied Sciences, Harvard University, Cambridge Massachusetts 02138, USA}

\author{Giuseppe Castaldi}
\affiliation{Department of Engineering, University of Sannio, 1-82100 Benevento, Italy}

\author{Massimo Moccia}
\affiliation{Department of Engineering, University of Sannio, 1-82100 Benevento, Italy}

\author{Vincenzo Galdi}
\affiliation{Department of Engineering, University of Sannio, 1-82100 Benevento, Italy}

\author{Shriram Ramanathan}
\affiliation{Harvard School of Engineering and Applied Sciences, Harvard University, Cambridge Massachusetts 02138, USA}

\author{Yuki Sato}
\email[]{sato@rowland.harvard.edu}
\affiliation{Rowland Institute at Harvard, Harvard University, Cambridge Massachusetts 02142, USA}

\date{\today}

\maketitle

\section{Abstract}

 We demonstrate an innovative multifunctional artificial material that combines exotic metamaterial properties and the environmentally responsive nature of phase change media. The tunable metamaterial is designed with the aid of two interwoven coordinate-transformation equations and implemented with a network of thin film resistors and vanadium dioxide ($VO_{2}$). The strong temperature dependence of $VO_{2}$ electrical conductivity results in a relevant modification of the resistor network behavior, and we provide experimental evidence for a reconfigurable metamaterial electric circuit (MMEC) that not only mimics a continuous medium but is also capable of responding to thermal stimulation through dynamic variation of its spatial anisotropy. Upon external temperature change the overall effective functionality of the material switches between a ``truncated-cloak" and ``concentrator" for electric currents. Possible applications may include adaptive matching resistor networks, multifunctional electronic devices, and equivalent artificial materials in the magnetic domain. Additionally, the proposed technology could also be relevant for thermal management of integrated circuits.

\section{Introduction}

Materials synthesis has reached a point where one can literally grow materials from scratch while manipulating the substance at the atomic and molecular level. This has led to various new possibilities for material properties that never existed in the past. A related field that has also seen a tremendous growth in recent years is metamaterial engineering. Artificial structures from nanoscales and up can be fabricated to collectively induce exotic properties not typically found in nature. Traditionally proposed as a platform for studying electromagnetic phenomena, metamaterials have expanded beyond the realm of optics \cite{wegener2013,maldovan2013} and seen applications in diverse areas including the manipulation of acoustic waves \cite{cummer2007}, magnetostatic fields \cite{narayana2012dc,magnus2008,gomory2012}, heat flux \cite{narayana2012,narayana2013,schittny2013} and electric current \cite{cuiprl2012} .

In the area where the fields of material science and metamaterial engineering overlap, the ability to reconfigure the response of the materials in real time has emerged as one of the most important goals.
For electromagnetic waves, the hybrid-metamaterial approach has been successful. Meta-structures with graphene \cite{papasimakis2010}, chalcogenide glass \cite{samson2010}, liquid crystals \cite{savo14}, NEMS/MEMS \cite{ou2011}, superconductors \cite{savinov2012} and vanadium dioxide (VO$_2$) \cite{driscoll2009,huang2010} have for instance resulted in linear and nonlinear control of transmission characteristics, yielding frequency tunability. In contrast to these time-varying counterparts, however, the incorporation of similar hybridized structures in static (or quasi-static) electric and magnetic metamaterials has been largely unexplored to date. Because of their intrinsically non-resonant nature, it is clear that a reconfigurable frequency response would not be the aim, but a more drastic operational shift may be feasible from a functional device made of DC hybrid-metamaterials.

In this manuscript we report the experimental integration of VO$_2$ phase change materials in DC electric metamaterial structures. We have designed a metamaterial electric circuit (MMEC) with cloaking-like and concentrating-like functionalities with the aid of a coordinate-transformation method theory \cite{pendry2006}. We have then implemented the MMEC on a 4-inch silicon wafer using thin film technology and circuit theory. The material switches by itself from one functionality to the other via VO$_2$ metal-to-insulator transitions induced by an external temperature change. The experimental results and the theoretical approach presented here demonstrate the concept of embedding functional substances in bigger artificial heterostructures to induce another level of functionalities, while pointing the way toward analogous investigations and potential applications in a multitude of electrical, magnetic, and thermal systems.

\section{Results}
We begin by viewing the material to be constructed as a system consisting of an ensemble of elements that collectively give rise to the material's overall characteristics. This conceptual transition from a continuous medium to an equivalent discretized version provides a powerful tool for designing the anisotropic properties of artificial materials and introduces a functional component better suited for practical implementations \cite{cuiprl2012,dede2013}. In this particular case, we mimic a continuous cylindrical conducting medium of conductivity $\bar{\bar\sigma}$ and thickness $t$ using a polar resistor network comprising of spokes and arcs of finite lengths $\Delta \rho$ and $\Delta \phi$, where $\rho$ and $\phi$ denote the radial and angular directions in cylindrical coordinates. The dimensions of spokes and arcs can be calculated from Ohm's law, $R_{\rho}=\Delta \rho/(t\sigma_{\rho}\rho \Delta \phi)$ and $R_{\phi}=\rho \Delta \phi/(t\sigma_{\phi} \Delta \rho)$. By properly assigning the dimensions and conductivities of the polar grid elements, it is possible to mimic the properties of any continuous medium and hence achieve unprecedented control over the electric current flowing through the material.

Needless to say, the resistive network topology itself is relevant for many circuit applications including filters \cite{zobel23,norton37}, matching networks for power electronics \cite{han07} and electric impedance tomography (EIT) \cite{borcea08,cheney1999}. Design strategies include minimization of power dissipation \cite{roos2012}, resistance distance \cite{klein93}, and dealing with inverse problems \cite{borcea08}. The choice strongly depends on the desired application. In this work we apply the resistive network to mimic a continuous medium while gaining control over the spatial anisotropy.

In tailoring spatial anisotropy for DC metamaterials, all previous works have focused on single functionalities (e.g. cloak, concentrator) \cite{cuiprl2012,cuiscirep2012,cuiapl2013} or made use of additional DC sources for altering the spatial distribution of electric currents \cite{cuiprl2013}. For realistic applications the ability of a metamaterial electric (and eventually electronic) circuits to readily respond to the surrounding environmental variations (e.g. temperature changes) is of strategic importance. Combining metamaterials with environmentally responsive material compounds could pave the way for such multifunctional metamaterial electric circuits.

The material discretization is followed by the assignment of resistance values for all constitutive elements. The multifunctional scenario considered here naturally requires two sets of such values with a way to switch between the two. To demonstrate reconfigurability, we have picked the examples of cloaking-like and concentrator-like functionalities for the two electric properties of the overall material, and we employ VO$_2$ as the temperature dependent switch. We first utilize the concept of coordinate-transformation to determine the spatial resistance profiles for the two scenarios. Such coordinate-transformation-based designs have the exclusive advantage of being naturally impedance matched \cite{pendry2006,schurig2006,leonhardt2006}, and for this reason the load variation arising from spatial resistance change will not be sensed by the external circuitry and sources. Furthermore, with regard to applications, designs based on coordinate-transformation could be employed to guarantee robustness against natural fluctuations of circuit parameters.

\begin{figure}
\centering
\includegraphics[width=4.5in]{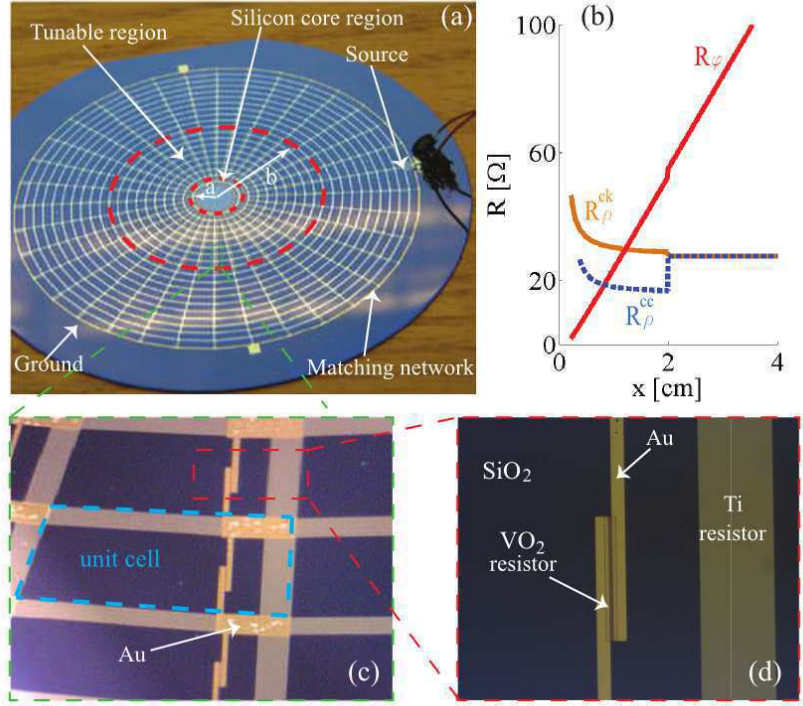}
\caption[]{(a) Camera image of the MMEC fabricated on a silicon wafer. The area between the red dashed circles is the tunable region. a and b are the dimensions of the core region and tunable region respectively. (b) Resistance values derived from coordinate-transformation equations (see equation (15)) used to calculate the dimensions of the titanium and $VO_{2}$ thin film resistors in the radial and angular directions R$_{\rho}^{ck}$, R$_{\rho}^{cc}$ and R$_{\phi}$. (c) Close up of the MMEC unit cell.(d) Close up of the $VO_{2}$/Au and Ti resistors. They are both deposited on a 100nm thermally grown $SiO_{2}$ layer (dark background region). }
\label{fig1}
\end{figure}

We refer readers to the appendix section for the details of the coordinate-transformation formalism. We note briefly here that, as the first type of anisotropic material, we induce the so-called ``truncated cloak". A well-established \cite{ruan2007} coordinate transformation of $f_{ck}(\rho)=b(\rho-a+\delta)/(b-a+\delta)$ is used to compress space from a cylindrical region $\rho<b$ into an annular region $a<\rho<b$ (see appendix A). As $\delta \rightarrow 0$, all electrical current would be removed from the core region $\rho<a$. For the second type of anisotropy, we induce a ``concentrator". For this case, instead of standard coordinate transformations used in the past \cite{cuiscirep2012}, we introduce a new transformation $f_{cc}(\rho)=b\left(f_{ck}(\rho)/b\right)^\gamma$ where $0<\gamma<1$. This new formalism effectively links the two transformations and makes the practical design for multifunctionality feasible. We plot in Fig. \ref{fig1}b the radial and angular components of the polar network resistance R$_{\rho}$ and R$_{\phi}$, obtained from the coordinate-transformation framework above for the two functionalities.

When R$_{\rho}$=R$_{\rho}^{ck}$ (orange continuous line), the overall material mimicked by the resistor grid functions as a truncated cloak. A perfect cloak would require infinite conductivity at the core boundary, making it prohibitive to simultaneously implement the anisotropy required for the concentrating behavior. For this reason we induce a truncated cloak as discussed above thereby allowing only a small fraction of the current to penetrate the inner region
while ensuring that the electrical current profile is not perturbed in the exterior region ($\rho>b$). When R$_{\rho}$=R$_{\rho}^{cc}$ (blue dashed line) the current is forced to flow into the core region, while the current profile remains the same in the exterior as the cloaking scenario. The degree of concentration depends nonlinearly on the geometric factors $a$ and $b$.

In Figure \ref{fig1}a, we show a camera image of our MMEC fabricated on a 4-inch silicon wafer. The resistance profiles R$_{\rho}^{ck}$ and R$_{\phi}$ are implemented by patterning titanium (Ti, $\rho_{Ti}=132\mu\Omega cm$) rectangles with fixed length ($1850\mu m$) and thickness ($200nm$) and variable width to match the desired resistance values at particular locations. To reconfigure the anisotropy of the overall material mimicked by the network, VO$_2$ thin film resistors have been grown in parallel with the Ti resistors oriented radially. The three orders of magnitude reduction in the VO$_2$ electric resistivity across the phase transition allows us to approximately match the lower radial resistance of R$_{\rho}^{cc}$ required. Figures \ref{fig1}c and \ref{fig1}d show closeups of the MMEC unit cells and the thin film patterns, respectively. Electrical links between resistors are guaranteed by rectangular Au pads positioned at every node. The center of the grid is left open and is electrically connected to the silicon wafer forming a core region with radius $a=2100\mu m$ and thickness $t=500\mu m$. The rest of the silicon wafer surface is made insulating with a layer of silicon dioxide. The dimensions, and hence resistances, of all the thin film elements on SiO$_2$ surrounding the core are obtained through coordinate-transformations with the core silicon ($\sigma=10 S/m$) as the ``background medium" taking its thickness into account to appropriately impedance-match it. Moreover the outermost circle acts as the ground and it is connected to the rest of the circuit through matching resistors made of tantalum nitride (TaN, $\rho_{TaN}=2280\mu\Omega cm$) to emulate an infinite material. Overall the network consists of a $20\times36$ polar array of thin film resistors and a $10\times36$ array of VO$_2$ elements. Details on the calculations regarding the thin films and matching resistors, dimensions and resistances of Au, Ti, TaNi, and VO$_2$ films, as well as detailed fabrication/growth steps are listed in the appendix C.

\begin{figure}
\centering
\includegraphics[width=6in]{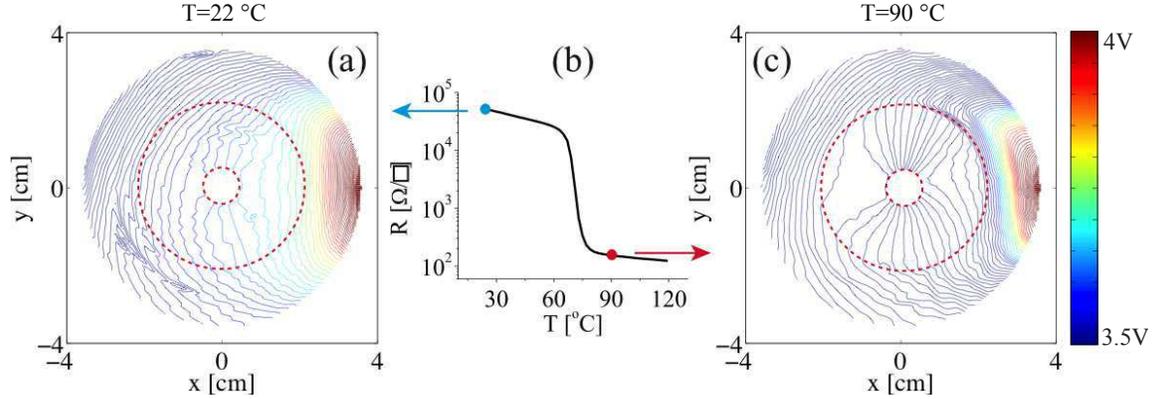}
\caption[]{Contour map of the voltage levels measured at each node of the MMEC. (a) Voltage map for the truncated cloak functionality measured at $22^{\circ}C$ and (c) voltage map for the concentrator functionality measured at $90^{\circ}C$. (b) Experimentally measured sheet resistance curve of the $VO_{2}$ used in our experiment. Blue and red points show the sheet resistance values at $22^{\circ}C$ and $90^{\circ}C$ respectively.}
\label{fig2}
\end{figure}

The characterization of the metamaterial electric circuit is carried out by mapping the voltage spatial distribution over the polar grid by probing the voltage at every node. The MMEC is fed with a $5V$ DC power supply (HP 3611A) connected to the $19^{th}$ node from the center of the polar grid (see fig. \ref{fig1}a). The voltage amplitude is measured with respect to the ground. A custom made probe station with three motorized linear axes is used to sense the voltage at each node. Two stepper motors are programmed to position the sample horizontally, whereas the third stepper motor moves a spring-loaded contact probe vertically. The electric signal is measured with a HP34401A multimeter. In order to induce the desired change of the VO$_2$ resistivity the sample was mounted on an electronically controlled hot chuck.

Figure \ref{fig2}a and \ref{fig2}c show the voltage maps measured at room temperature of $22^{\circ}C$ and at $90^{\circ}C$, respectively. The region within the two red dashed circles is the tunable region. In fig. \ref{fig2}b we also plot the VO$_2$ sheet resistance as a function of the temperature. The metal-insulator transition for the VO$_2$ used in this experiment was designed to take place between $70^{\circ}C$ and $75^{\circ}C$. The measured voltage contour maps are markedly different at the two operating temperatures. The voltage distribution at T=$22^{\circ}C$ shows the typical signature of a truncated cloak \cite{ruan2007}, where external equipotential contour lines distribute almost undisturbed as if the tunable metamaterial region did not exist. Only in proximity of the outermost red dashed circle the equipotential lines show some perturbation. This is in agreement with the coordinate-transformation equations used for the design of our MMEC (see appendix A).  At T=$90^{\circ}C$ the profile changes drastically (see Figure \ref{fig2}c) as a result of the VO$_2$ insulator-to-metal transition and shows a voltage distribution characteristic of a concentrator. The contour lines are pulled towards the core region while the exterior region shows minimum perturbation. Under this condition the VO$_2$ behaves as a conductor and the equivalent resistance of the radial branches is significantly lower.

\begin{figure}
\centering
\includegraphics[width=4in]{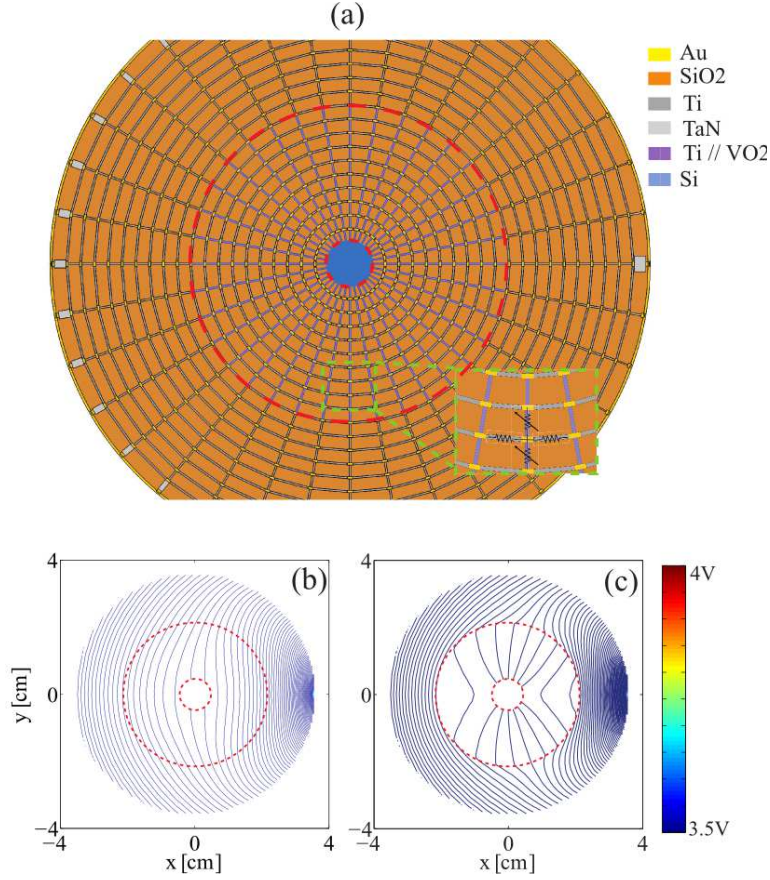}
\caption[]{(a) Color coded COMSOL model used for the simulation of the MMEC response. The inset shows a closeup of the tunable region where thin film resistors are modeled with rectangular elements oriented in the radial (variable resistor symbol) and angular directions (fixed resistor symbol). The anisotropic properties are changed by varying the resistivity of the purple elements. (b) Numerically calculated voltage map for the truncated cloak functionality and (c) for the concentrator functionality. The source is place to the right side, where the density of the equivoltage lines is higher. The area enclosed between the red dashed circles is the tunable region.}
\label{fig3}
\end{figure}

Our experimental results are confirmed by finite-element numerical simulations performed via the commercial software package COMSOL \cite{comsol}. Figure \ref{fig3}a shows a color coded schematics of the 2D model used to simulate the MMEC. Each color refers to a different material used to implement the model. The reader can refer to the appendix C for the electric conductivity values used to describe the model. The conductivity of the silicon inner region is made 2500 times larger in this 2D simulation in order to account for the thickness difference between the $200nm$ conductive thin films forming the polar grid and the silicon core which is $500\mu m$ thick. The resistors R$_{\rho}$ are modeled through resistive elements having the same pattern as the titanium thin films. The transition from R$_{\rho}^{ck}$ to R$_{\rho}^{cc}$ is obtained by changing the conductivity of purple elements located inside the tunable region, enclosed between the two concentric red dashed circles (see inset Figure \ref{fig3}a). The truncated-cloaking and concentrator functionalities measured experimentally are matched well by the numerical simulations when the conductivity of the purple spokes is varied by a factor of 14. Based on our experimental implementation of Au/VO$_2$ bridges in parallel with Ti, we expect the resistance in the radial direction to change by a factor of 18 at the VO$_2$ phase transition. The two corresponding maps with the numerical results are shown in Figure \ref{fig3}b and Figure \ref{fig3}c.

The amplitude differences between the experimental and numerical contour maps are primarily caused by the conductivities of the TaN matching resistors next to the grounding circle. TaN is deposited by means of sputtering, and its final conductivity is strongly dependent on the N$_{2}$ flow rate. We have found that this dependence is highly nonlinear especially in proximity of the target values, making it nontrivial to meet the exact conductivity values required. Subsequent tests of the TaN resistivity values have confirmed a two fold difference with respect to the expected values. The present design is very demanding in terms of fabrication tolerances because the differences between the dimensions of neighboring resistors are small, ranging between $1\mu m$ and $90\mu m$. Typical tolerances of the contact aligner ($+/-0.5\mu m$) for example could affect the above dimensions enough to cause noticeable fluctuations in the voltage profiles. This was made even more challenging as the MMEC required patterning of very small features over a large surface. The turbulent profile of some of the contour lines may be related to this non-uniformity of thin film resistors across the entire 4-inch wafer. Despite the aforementioned matters, the experimental results clearly demonstrate the reconfigurable anisotropy of the metamaterial.

The experimental results reported in this manuscript suggest that the metamaterial paradigm may be used to provide solutions for tangible compact electronics using thin film technologies. Traditionally, electronic systems comprise of several blocks, each designed to play a specific role. A smarter design approach could rely on a single unit capable of handling multiple functions such as carrying electric current and heat simultaneously but independently. A potential first step in this direction has been shown recently using metamaterial structures designed with the framework of multiphysics transformations \cite{moccia2013}. The same goal may be approached through a different path of utilizing phase change materials that have varying responses in different physical domains.

For example it has been shown that VO$_2$ exhibits a change in its thermal conductivity upon phase transition, but it is dwarfed by the variation in its electrical conductivity \cite{andreev1978,chen2012,oh2010}. This asymmetry in the two physical domains may be utilized to obtain two independent functionalities for heat and electricity. One can imagine adding right on top of a MMEC an additional network of material (such as BeO$_2$) with high thermal conductivity and limited electrical conductivity. When the metal-to-insulator transition takes place, only the electrical current profile within the overall material would be reconfigured, leaving the heat flux distribution unchanged. It is also conceivable to design layers to reconfigure both kinds of transport simultaneously but in different fashions. This line of bilayer approach, which takes advantage of nonlinear phase change materials with strongly different electrical and thermal properties, may be useful for designing more compact electronic components with nontrivial multifunctionalities.
We note in relation to this that recently a beautiful experimental work has been reported where thin film technology combined with coordinate-transformation approach are used to fabricate systems for the unconventional manipulation of heat flux using the discretization of thermal conductors to introduce anisotropy \cite{dede2013}. Potential applications include thermal management of integrated circuits such as responding to temperature changes and rerouting electrical current away from hot spots while reconfiguring its path in a prescribed and pseudo-automated fashion.

\section{Conclusions}

In conclusion, this work presents the development and characterization of a reconfigurable metamaterial electric circuit, designed to respond to temperature changes through dynamic variation in its spatial anisotropy. The tunable metamaterial device is built by combining a network of thin film resistors and vanadium dioxide (VO$_2$). The strong temperature dependence of the VO$_2$ electrical conductivity results in a significant modification of the resistor network behavior, giving rise to a new level of functionalities. Through a new set of coordinate-transformation equations, used to tailor the response of the metamaterial circuit upon temperature change, we have shown that our MMEC mimicking a continuous and anisotropic medium can act as a truncated cloak or concentrator for electric currents. When designing an electronic system, the influence of the surrounding environment can be considered either a nuisance and hence something to be suppressed or a convenient knob that can be turned to affect the material properties. In the latter view, an ideal scenario would be to have the system designed in such a way that the external effect could be amplified and not only the local material characteristics but also the collective device property can be drastically manipulated with environmental stimuli. For such device designs and fabrications, a new paradigm could appear from the marriage of environmentally responsive materials - phase change materials - and the framework of metamaterials tailored with the mathematical formalism of coordinate-transformation.
\section{Acknowledgments}
S. S. and Y. S. acknowledge support from the Rowland Institute at Harvard University. S.R. acknowledge NSF for financial support.

\section{Appendix A: Coordinate-Transformation Method}
\begin{figure}
\begin{center}
\includegraphics [width=16cm]{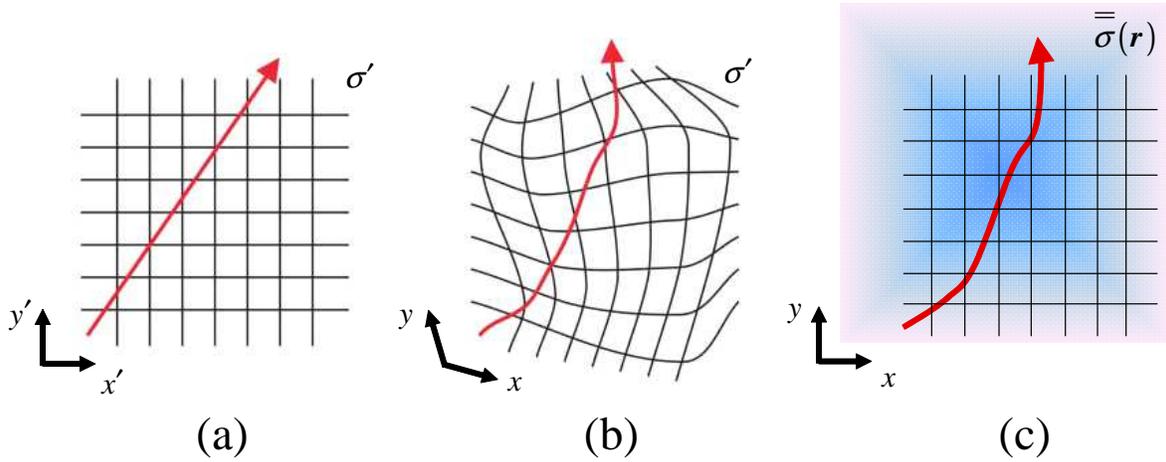}
\end{center}
\caption{ Path of the electric current density (red arrow) in (a) a Cartesian coordinates space filled with an isotropic material, (b) transformed coordinates space filled with an isotropic material, (c) Cartesian coordinate space filled with inhomogeneous and anisotropic transformation medium with conductivity tensor ${\bar {\bar \sigma}}$. }
\label{FigureSM1}
\end{figure}

We start considering an auxiliary space ${\bf r}'\equiv(x',y',z')$ filled with an isotropic material with electrical conductivity $\sigma'$ [cf. Figure \ref{FigureSM1}(a)]. At equilibrium, the sourceless electrical conduction equation is given by
\beq
\nabla
\cdot\left[\sigma'\left({\bf r}'\right)\nabla V'\left({\bf r}'\right)\right]=0,
\label{eq:ECC}
\eeq
with $V'$ denoting the electrical potential. Assuming, for instance, a homogeneous conductivity distribution $\sigma'$, and applying a constant voltage along a given direction, the electrical current density follows a straight path. Next, we apply a coordinate transformation,
\beq
{\bf r}'={\bf f}\left({\bf r}\right),
\label{eq:CT}
\eeq
to a {\em curved-coordinate} space ${\bf r}$, which modifies the path of the electrical current density in a desired fashion [cf. Figure \ref{FigureSM1}(b)]. In view of the co-variance properties of equation (\ref{eq:ECC}), the transformation induced in the electric potential in this new space \cite{pendry2006},
\beq
V\left({\bf r}\right)=V'\left[{\bf f}\left({\bf r}\right)\right],
\eeq
can be equivalently obtained in a {\em flat} Cartesian space ${\bf r}\equiv(x,y,z)$  filled with an {\em inhomogeneous}, {\em anisotropic} ``transformation medium'' [cf. Figure \ref{FigureSM1}(c)] characterized by a conductivity tensor \cite{pendry2006}
\beq
{\bar {\bar \sigma}}\left({\bf r}\right)=\sigma'
\det\left({\bar {\bar \Lambda}}\right){\bar {\bar \Lambda}}^{-1}\cdot
{\bar {\bar \Lambda}}^{-T},
\label{eq:sigma}
\eeq
where
\beq
{\bar {\bar \Lambda}}=\frac{\partial {\bf f}}{\partial {\bf r}}
\eeq
is the Jacobian matrix associated with the coordinate transformation in equation (\ref{eq:CT}), and the superscripts ``-1'' and ``-T'' denote the inverse and inverse-transpose, respectively.

In particular, by considering a {\em radial} transformation,
\beq
\rho'=f\left(\rho\right),
\eeq
between the associated cylindrical systems ($\rho',\phi',z'$) and ($\rho,\phi,z$), the relevant components of the conductivity tensor in equation (\ref{eq:sigma}) can be written as
\beq
\sigma_{\rho}\left(\rho\right)=\sigma' \frac{f\left(\rho\right)}{\rho {\dot f}\left(\rho\right)},~~
\sigma_{\phi}\left(\rho\right)=\sigma' \frac{\rho {\dot f}\left(\rho\right)}{f\left(\rho\right)}.
\label{eq:sigmarp}
\eeq
with the overdot denoting differentiation with respect to the argument.

\begin{figure}
\begin{center}
\includegraphics [width=16cm]{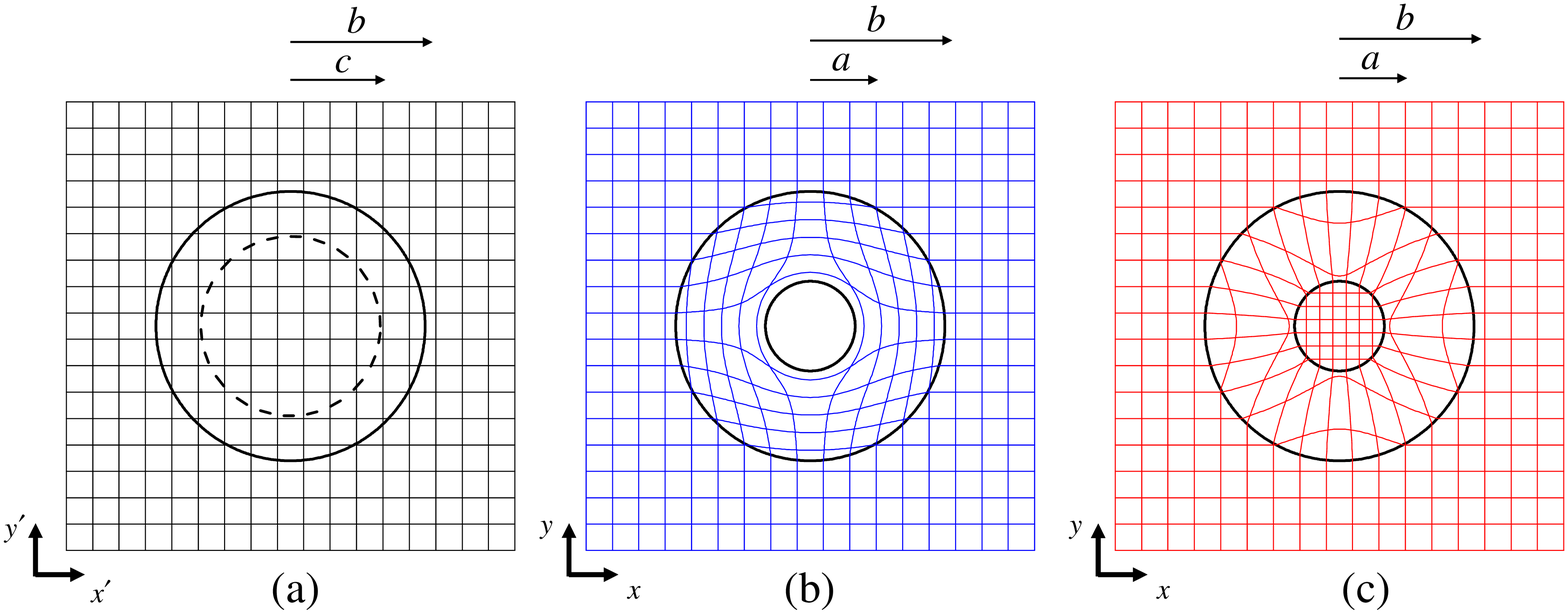}
\end{center}
\caption{Schematic illustration of the coordinate transformations. (a) Auxiliary space; (b) and (c) Transformed spaces for the cloak and concentrator transformations, respectively.}
\label{FigureSM2}
\end{figure}

In our study, we are interested in switching between two functionalities induced by different coordinate transformations. More specifically,  we consider an ``invisibility-cloak'' functionality induced by the coordinate transformation
\beq
f_{ck}\left(\rho\right)=\frac{b\left(\rho-a+\delta\right)}{b-a+\delta},
\label{eq:fck}
\eeq
and a ``concentrator'' functionality induced by the coordinate transformation
\beq
f_{cc}\left(\rho\right)=b\left[
\frac{f_{ck}\left(\rho\right)}{b}
\right]^\gamma,~~0<\gamma<1.
\label{eq:fcc}
\eeq
As illustrated in Figure \ref{FigureSM2}, for small values of $\delta$, the above cloak and concentrator transformations essentially map
an annular cylinder of radii $a$ and $b>a$ in the transformed space ${\bf r}$ onto a cylinder of radius $b$ and an annular cylinder of radii $c>a$ and $b$, respectively, in the auxiliary space  ${\bf r}'$. Figure \ref{FigureSM2} also illustrates the coordinate distortions induced by these two transformations.

We highlight that the transformations in equations (\ref{eq:fck}) and (\ref{eq:fcc}) are {\em intertwined}, and they are slightly different from the ones conventionally used in the literature \cite{cuiprl2012,rahm}. In particular, the transformation in equation (\ref{eq:fck})
actually yields an {\em imperfect} cloaking effect (with the ideal case recovered in the limit $\delta\rightarrow 0$), while the transformation in equation (\ref{eq:fcc}) is purposely chosen so as to simplify the subsequent implementation.

Via straightforward application of equation (\ref{eq:sigmarp}), we can analytically derive the expressions of the conductivity profiles associated with the two transformations. More specifically, for the cloak functionality, we obtain
\beq
\sigma_{\rho}^{ck}\left(\rho\right)=\sigma' \left(\frac{\rho-a+\delta}{\rho}\right),~~
\sigma_{\phi}^{ck}\left(\rho\right)=\sigma' \left(\frac{\rho}{\rho-a+\delta}\right),
\label{eq:sigmac1}
\eeq
from which we observe that the approximate character of the transformation in equation (\ref{eq:fck}), for {\em finite} values of $\delta$, prevents the arising conductivity components to exhibit {\em extreme} values at the inner boundary $\rho=a$, viz.,
\beq
\sigma_{\rho}^{ck}\left(a\right)= \frac{\sigma' \delta}{a},~~
\sigma_{\phi}^{ck}\left(a\right)= \frac{a}{\sigma' \delta}.
\label{eq:sigma_ck}
\eeq
As a consequence, an {\em imperfect} cloaking effect is attained, with the electrical current density inside the inner region $\rho<a$ not exactly vanishing, but reduced by a factor proportional to
\beq
\chi_{ck}=\frac{\delta}{a}.
\label{eq:chi_ck}
\eeq
On the other hand, for the concentrator functionality, we obtain
\beq
\sigma_{\rho}^{cc}\left(\rho\right)=\frac{\sigma_{\rho}^{ck}}{\gamma}=\sigma' \left(\frac{\rho-a+\delta}{\gamma \rho}\right),~~
\sigma_{\phi}^{cc}\left(\rho\right)=\gamma \sigma_{\phi}^{ck}\left(\rho\right) =\gamma \sigma' \left(\frac{\rho}{\rho-a+\delta}\right).
\label{eq:sigma_cc}
\eeq
In this case, the electrical current density inside the inner region $\rho<a$ is enhanced by a factor
\beq
\chi_{cc}=\frac{f_{cc}\left(a\right)}{a}=\frac{b}{a}\left(
\frac{\delta}{b-a+\delta}
\right)^{\gamma}>1.
\label{eq:chi_cc}
\eeq

The parameters $\delta$ and $\gamma$ can be computed from equation (\ref{eq:chi_ck}) and (\ref{eq:chi_cc}), once the desired cloaking and concentration factors $\chi_{ck}$ and $\chi_{cc}$ have been set. In our design, we chose $\delta=0.0021m$ and $\gamma=0.7613$.

We note from equations (\ref{eq:sigma_ck}) and (\ref{eq:sigma_cc}) that the aforementioned {\em intertwining} of the two transformations yields, for the two functionalities, conductivity components that differ only by multiplicative constants. This significantly simplifies the subsequent implementation, based on a polar resistor network comprising of spokes and arcs of finite lengths $\Delta\rho$ and $\Delta\phi$, respectively, and thickness $t$. The resistance profiles $R_{\rho}$ and $R_{\phi}$ in this network are readily related to the conductivity profiles via Ohm's law \cite{cuiprl2012},
\beq
R_{\rho}\left(\rho\right)=\frac{\Delta\rho}{\sigma_{\rho} \rho t \Delta\phi},~~
R_{\phi}\left(\rho\right)=\frac{\rho \Delta\phi}{\sigma_{\phi} t \Delta\rho}.
\label{eq:RR}
\eeq
From equation (\ref{eq:RR}), it is evident that the switching between the cloak and concentrator functionalities requires the reconfiguration of {\em both} resistance profiles $R_{\rho}$ and $R_{\phi}$. However, in our proposed design, we followed a simplified approach by assuming
\beq
R_{\phi}^{ck}\left(\rho\right)=R_{\phi}^{cc}\left(\rho\right),
\label{eq:RR1}
\eeq
so that only the radial resistance profile $R_{\rho}$ need to be reconfigured. In order to better understand the approximation underlying this assumption, it is insightful to look at the conductivity profiles generated by assuming that the cloak and concentrator transformations are applied to auxiliary spaces with different background conductivities $\sigma'_{ck}$ and $\sigma'_{cc}$, respectively, viz.,
\beq
\sigma_{\rho}^{ck}\left(\rho\right)=\sigma'_{ck} \left(\frac{\rho-a+\delta}{\rho}\right),~~
\sigma_{\phi}^{ck}\left(\rho\right)=\sigma'_{ck} \left(\frac{\rho}{\rho-a+\delta}\right),
\eeq
\beq
\sigma_{\rho}^{cc}\left(\rho\right)=\sigma'_{cc} \left(\frac{\rho-a+\delta}{\gamma \rho}\right),~~
\sigma_{\phi}^{cc}\left(\rho\right)=\gamma \sigma'_{cc} \left(\frac{\rho}{\rho-a+\delta}\right).
\label{eq:sigma1}
\eeq
We note from equation (\ref{eq:RR}) that the simplifying assumption in equation(\ref{eq:RR1}) implies
\beq
\sigma_{\phi}^{ck}\left(\rho\right)=\sigma_{\phi}^{cc}\left(\rho\right),
\eeq
which can be achieved from equation (\ref{eq:sigma1}) by assuming
\beq
\sigma'_{ck}=\gamma \sigma'_{cc},
\eeq
and also yields
\beq
\frac{R_{\rho}^{ck}\left(\rho\right)}{R_{\rho}^{cc}\left(\rho\right)}=
\frac{\sigma_{\rho}^{cc}\left(\rho\right)}{\sigma_{\rho}^{ck}\left(\rho\right)}=\gamma^2.
\label{eq:RR2}
\eeq
We observe from equation (\ref{eq:RR2}) that, in view of the judicious intertwining of the two transformations, the reconfiguration required on the radial resistance profile amounts only to a constant scaling factor.

The above interpretation, in terms of cloak and concentrator transformations applied to two different and properly chosen auxiliary spaces, inherently implies that it is no longer possible to achieve perfect impedance matching for both functionalities. In our proposed design, we chose to enforce the perfect impedance matching for the cloak functionality, thereby allowing a moderate mismatch on the concentrator side.

%

%


\section{Appendix B: $VO_{2}$ Deposition Method}

\begin{figure}
\begin{center}
\includegraphics [width=10cm]{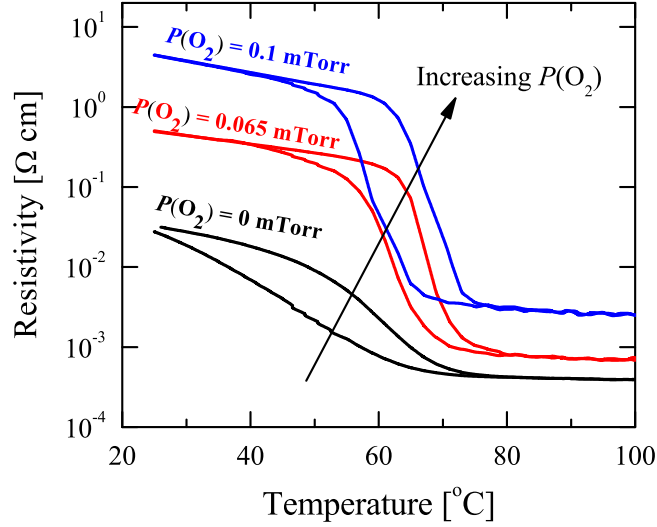}
\end{center}
\caption{Plot of the resistivity measured for increasing levels of $O_{2}$ partial pressure. Experimental results were carried out on a 100nm thick $VO_{2}$ thin film.}
\label{FigureSM3}
\end{figure}

 $VO_{2}$ thin films were grown on 100 nm thick thermal $SiO_{2}$ on Si wafer by RF magnetron sputtering from a $V_{2}O_{5}$ target in an $Ar/O_{2}$ gas mixture. The growth temperature and total pressure were kept constant at $550^{\circ}C$ and 5 mTorr during the deposition, respectively. It is known that the resistivity and the metal-insulator transition temperature of $VO_2$ are extremely sensitive to any subtle change in the oxygen partial pressure during the growth. As a result, the oxygen partial pressure was carefully controlled by adjusting the relative flow rate of pure Ar and that of $10\%$ $O_{2}$ balance Ar, which allows us to tune the oxygen partial pressure down to 0.02 mTorr when the total pressure is fixed at 5 mTorr. Figure \ref{FigureSM3} shows the resistivity-temperature curves of $VO_2$ samples grown under different oxygen partial pressure. The resistivity was measured by 4-probe Van der Pauw measurements on a temperature-controlled hot chuck. The thickness of the films and therefore growth rate were calibrated by X-ray reflectivity and optical ellipsometry. The resistivity of $VO_2$ in both the insulating and metallic phases decreases with decreasing oxygen partial pressure, because of the extra free electrons donated by oxygen vacancies formed during the growth. When the oxygen partial pressure is zero, both the metal-insulator transition magnitude and temperature are suppressed by oxygen vacancy formation. On the other hand, if the oxygen partial pressure exceeds 0.1 mTorr, the deposited films are further oxidized into $V_{2}O_{5}$ that does not exhibit any metal-insulator transitions (not shown). As a result, in order to achieve large on/off ratio and have reasonable device dimensions, for the present work we chose the resistivity-temperature curve of samples grown under oxygen partial pressure of 0.065 mTorr.

%


\section{Appendix C: Device Parameters}

The MMEC design comprises of two steps. During the first step we calculate the resistance values $R_{\rho}$ and $R_{\phi}$ from equation (\ref{eq:RR}). To validate our design we run numerical simulations of the polar grid network using COMSOL. For the device fabricated experimentally with thin film technologies, different resistance values are achieved by using materials of fixed conductivities but varying the dimensions of relevant spokes and arcs making up the discretized grid. For numerical simulations, to simplify the computation, spokes and arcs of fixed dimensions are used while varying their electrical conductivities in order to match the desired anisotropic profile obtained from equation (\ref{eq:RR}). The second design step involves the calculation of the dimensions of the polar grid for the real device. Table \ref{tab1} lists the values of $R_{\rho}^{ck}$ and $R_{\phi}$ of the thin film resistors used in our experiment and the corresponding values for their widths. Lengths and angular dimensions are fixed and their values are $\Delta\rho = 1850\mu m$ and $\Delta\phi=10^{\circ}$. Resistors have been fabricated with Titanium films ($\rho_{Ti}=132\mu\Omega cm$) and the film thickness is fixed at $t_{Ti}=200nm$.

\begin{figure}
\begin{center}
\includegraphics [width=6cm]{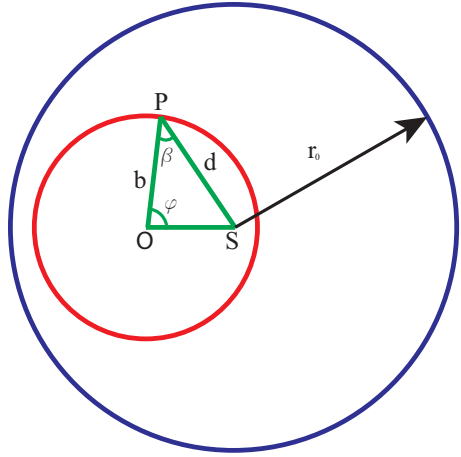}
\end{center}
\caption{Schematic showing the geometrical parameters used for the derivation of the matching resistors network}
\label{FigureSM4}
\end{figure}

In table \ref{tab1} are also listed the resistances $R_{m}$ and widths $w_{Rm}$ of the matching network. The values are derived from
\beq
R_{m}=\frac{d(\log r_{0}-\log d)}{\sigma_{0} b \Delta\phi h cos\beta}
\label{eq:Rm}
\eeq

to emulate an infinite medium \cite{cuiprl2012}. In Figure \ref{FigureSM4} are shown all the variables of the last equation. In particular $r_{0}$ is the distance between the ground and the source point S, $\sigma_{0}$ is the conductivity of the background medium ({$\sigma_{0}=600S/m$}), $h=0.5mm$ and $\Delta\phi=10^{\circ}$. $R_{m}$ resistors are made with TaN ($\rho_{TaN}=2280\mu\Omega cm$). The $w_{Rm}$ values are calculated through Ohm's law where we fix the radial length ($\Delta\rho = 1850\mu m$) and the thickness ($t_{TaN}=200nm$). Although for our experiment we utilize the central core region of a Silicon wafer ($\sigma_{Si}=10S/m$) as the ``background medium", we set $\sigma_0=60\sigma_{Si}$ here. This allows us to design the surrounding material (emulated by the network grid) that is slightly more conducting than the core medium, which ensures stronger cloaking properties.

In Table \ref{tab2} are listed the resistance values $R_{VO_{2}}^{insul}$ and $R_{VO_{2}}^{cond}$ of the $VO_{2}$ in the insulating ($T=22^{\circ}$) and conductive state ($T=90^{\circ}$). Through Ohm's law we calculate the corresponding widths of the resistors $w_{VO_{2}}$. Also for these $VO_2$ resistors we have fixed the lengths ($l_{VO_{2}}= 15\mu m$) and thicknesses ($t_{VO_{2}}= 300nm$).

%


\begin{table}[t]
\begin{minipage}[b]{0.45\linewidth}\centering
\begin{tabular}{|c|c|c|c|c|}
      \hline
n & $R_{\rho}^{ck} [\Omega]$ & $R_{\phi} [\Omega]$ & $w_{\rho} [\mu m]$& $w_{\phi} [\mu m]$ \\ \hline
1 &    39.26 &     3.79 &   217.93 &   228.11 \\
2 &    33.75 &     8.80 &   253.47 &   314.72 \\
3 &    31.77 &    13.99 &   269.31 &   334.11 \\
4 &    30.73 &    19.28 &   278.36 &   341.17 \\
5 &    30.10 &    24.61 &   284.19 &   344.70 \\
6 &    29.68 &    29.93 &   288.24 &   347.11 \\
7 &    29.37 &    35.29 &   291.24 &   348.38 \\
8 &    29.14 &    40.66 &   293.55 &   349.25 \\
9 &    28.96 &    46.00 &   295.36 &   350.11 \\
10 &    28.82 &    51.37 &   296.85 &   350.56 \\
11 &    27.50 &    59.24 &   311.09 &   336.19 \\
12 &    27.50 &    64.62 &   311.09 &   337.67 \\
13 &    27.50 &    70.00 &   311.09 &   338.92 \\
14 &    27.50 &    75.38 &   311.09 &   340.00 \\
15 &    27.50 &    80.76 &   311.09 &   340.93 \\
16 &    27.50 &    86.14 &   311.09 &   341.75 \\
17 &    27.50 &    91.53 &   311.09 &   342.47 \\
18 &    27.50 &    96.91 &   311.09 &   343.11 \\
19 &    27.50 &   102.29 &   311.09 &   343.68 \\ \hline
 \hline
\end{tabular}
\end{minipage}
\hspace{0.5cm}
\begin{minipage}[b]{0.45\linewidth}
\centering
\begin{tabular}{|c|c|c|c|c|c|}
           \hline
n & $R_{m}$  &   $w_{R_{m}}$ & n & $R_{m}$  &   $w_{R_{m}}$\\ \hline
1 &   100.30 &  2313.13 & 20 &   100.30 &  2313.13 \\
2 &  1032.63 &   224.67 & 21 &  1032.63 &   224.67 \\
3 &  1580.63 &   146.78 & 22 &  1580.63 &   146.78 \\
4 &  1577.18 &   147.10 & 23 &  1577.18 &   147.10 \\
5 &  1416.57 &   163.78 & 24 &  1416.57 &   163.78 \\
6 &  1231.04 &   188.46 & 25 &  1231.04 &   188.46 \\
7 &  1056.33 &   219.63 & 26 &  1056.33 &   219.63 \\
8 &   900.91 &   257.52 & 27 &   900.91 &   257.52 \\
9 &   765.49 &   303.08 & 28 &   765.49 &   303.08 \\
10 &   648.67 &   357.66 & 29 &   648.67 &   357.66 \\
11 &   548.68 &   422.84 & 30 &   548.68 &   422.84 \\
12 &   463.85 &   500.16 & 31 &   463.85 &   500.16 \\
13 &   392.78 &   590.66 & 32 &   392.78 &   590.66 \\
14 &   334.34 &   693.90 & 33 &   334.34 &   693.90 \\
15 &   287.63 &   806.59 & 34 &   287.63 &   806.59 \\
16 &   251.96 &   920.78 & 35 &   251.96 &   920.78 \\
17 &   226.83 &  1022.81 & 36 &   226.83 &  1022.81 \\
18 &   211.88 &  1094.97 & 37 &   211.88 &  1094.97 \\ \hline
19 &   206.92 &  1121.22  & & &\\ \hline
       \hline
\end{tabular}
\end{minipage}
\caption[]{(Left) Table with the resistance values $R_{\rho}^{ck}$ and $R_{\phi}$ used to design the resistor network and the corresponding dimensions of the thin film widths $w_{\rho}$ and $w_{\phi}$. They are sorted, through the index n, from closest to farthest to the center of the grid. (Right) Table with the resistance values $R_{m}$ used to design the matching network and the corresponding dimensions of the thin film widths $w_{R_{m}}$. They are sorted, through the index n, in terms of angular position starting from the element aligned with the source along the radial direction and moving counterclockwise}
\label{tab1}
\end{table}

\begin{table}[t]
\begin{minipage}[b]{0.45\linewidth}\centering
\begin{tabular}{|c|c|c|c|}
\hline
 n & $R_{VO_{2}}^{insul} [\Omega]$ & $R_{VO_{2}}^{cond} [\Omega]$ & $w_{{VO_{2}}}[\mu m]$ \\ \hline
 1 &   715.29  &      2.15  &   1615.23 \\
 2 &   614.99  &      1.85  &   1878.64 \\
 3 &   578.81  &      1.74  &   1996.10 \\
 4 &   559.99  &      1.69  &   2063.18 \\
 5 &   548.50  &      1.65  &   2106.38 \\
 6 &   540.80  &      1.63  &   2136.37 \\
 7 &   535.23  &      1.61  &   2158.63 \\
 8 &   531.02  &      1.60  &   2175.71 \\
 9 &   527.76  &      1.59  &   2189.18 \\
 10 &   525.12  &      1.58  &   2200.17 \\ \hline
\hline
\end{tabular}
\end{minipage}
\caption[]{Table with the resistance values  $R_{VO_{2}}^{insul}$ and  $R_{VO_{2}}^{cond}$ used to design the $VO_{2}$ resistors and the corresponding dimensions of the thin film widths $w_{{VO_{2}}}$. They are sorted, through the index n, from closest to farthest to the center of the grid.}
\label{tab2}
\end{table}

\clearpage


\bibliographystyle{ieeetr}
\bibliography{bib2}

\end{document}